\documentclass[aps,prev,twocolumn,preprintnumbers,floatfix,nofootinbib]{revtex4-1}
\pdfoutput=1 

\usepackage{amsmath}
\usepackage{amsfonts}
\usepackage{amssymb}
\usepackage{xcolor}
\usepackage{graphicx}
\usepackage{times}
\usepackage{hyperref}

\begin{document}

\preprint{IIPDM-2019}

\title{\boldmath Has AMS-02 Observed  Two-Component Dark Matter?}

\author{Stefano Profumo}
\email{profumo@ucsc.edu}
\affiliation{Department of Physics and Santa Cruz Institute for Particle Physics, University of California, Santa Cruz, CA 95064, USA}
\author{Farinaldo Queiroz}
\email{farinaldo.queiroz@iip.ufrn.br}
\author{Clarissa Siqueira}
\email{csiqueira@iip.ufrn.br}
\affiliation{International Institute of Physics, Universidade Federal do Rio Grande do Norte,
Campus Universit\'ario, Lagoa Nova, Natal-RN 59078-970, Brazil}

\begin{abstract}
There is convincing observational evidence for an increasing cosmic-ray positron-to-electron ratio at energies larger than $\sim 10$~GeV, at odds with expectations from secondary positron production. The most recent AMS-02 data exhibit an interesting spectral feature consisting of a bump at an energy around $300$~GeV followed by a drop around $\sim 800$~GeV. A possible explanation to the most recent data is that the excess positron originates from decaying dark matter. Here, we show that models consisting of two dark matter particle species contributing equally to the global  cosmological dark matter density provide  strikingly good fits to the data. The favored models, with a best-fit with $\chi^2/d.o.f \sim 0.5$ consist of a first species weighing $750$~GeV decaying with a lifetime $\tau_{\chi}\sim 10^{26}$~s to $\tau$ lepton pairs  (or to a pair of vector bosons subsequently decaying to a $\tau$ pair each), and a second species with a mass around 2.3 TeV decaying to $\mu$ lepton pairs. We provide a few possible concrete realizations for this scenario. 

\end{abstract}

\maketitle

\section{Introduction}
\label{sec:intro}

Cosmic-ray 
measurements have improved our understanding of high-energy processes over the past decades since they are sensitive to both the nature of  particles produced in astrophysical phenomena and to the intergalactic medium via the diffusion and energy loss processes they undergo in the Galaxy. The Payload for Antimatter Matter Exploration and Light-nuclei Astrophysics (PAMELA) \cite{Adriani:2008zr} first provided convincing, highly statistically significant evidence of a rise in the cosmic-ray positron fraction at energies above $\sim10$~GeV. This feature was later confirmed by Fermi-LAT  employing the Earth shadow and geomagnetic field to discriminate the charge of the cosmic-rays \cite{FermiLAT:2011ab}. In 2013 the AMS-02 collaboration  confirmed the rise in the positron fraction with much better statistics, and extending the measurement up to $350$~GeV \cite{Aguilar:2013qda}, finding a relatively flat positron fraction for energies above $150$~GeV.

Several attempts to explain the rise in the positron fraction have been put forth based on a new injection source of primary positrons. The proposals rely either on dark matter annihilation, with a pair-annihilation cross section much larger than the thermal value, $3\times 10^{-26} \mathrm{cm}^3/\mathrm{s}$ \cite{Cholis:2013psa,Ibarra:2013zia}, thus strongly disfavored by independent gamma-ray observations in the directions of dwarf spheroidal galaxies \cite{Ackermann:2015zua}, or on decaying dark matter with a lifetime of the order of $10^{27}$~s \cite{Nardi:2008ix,Arvanitaki:2008hq}, or on the presence of nearby astrophysical objects \cite{Profumo:2008ms, Hooper:2008kg,Grasso:2009ma}.  Intriguingly, recent observations made by the High-Altitude Water Cherenkov Observatory (HAWC) have confirmed the presence of energetic electrons and positrons from nearby pulsars, but the inferred diffusion parameter would rule out nearby, mature pulsars such as Geminga and Monogem as the sources of the rising positron fraction \cite{Abeysekara:2017old}, unless pulsars reside inside inefficient diffusion regions \cite{Profumo:2018fmz}. It is fair to state that the origin of the positron excess is at present unknown.

New AMS-02 measurements \cite{Aguilar:2019owu} revealed interesting new features. First, the observations confirmed the rise in the positron fraction for energies above $10$~GeV, with even greater statistics. Second, the previous flat differential spectrum (times energy to the third power) for energies larger than $150$~GeV was found to have a bump-like feature with a peak around $300$~GeV. Third, a cut-off at energies around $800$~GeV is now visible. We deem it timely to re-analyze this significantly improved spectral measurement. As we show here, AMS-02 might potentially show evidence for a two-component, decaying dark matter scenario.

A critical aspect of our study is which spectrum to assume for the background flux of positrons of secondary origin, i.e. originating from an inelastic collision of primary cosmic rays with the interstellar medium. Here, we adopt the background recommended by the AMS collaboration, so our conclusions rely on its validity. A departure from the background recommended by the collaboration would affect our conclusions, together with different assumptions on  propagation and energy losses \cite{Aguilar:2019owu}. 

In this work, we interpret the positron excess in terms of decaying dark matter. There exist stringent bounds on the dark matter lifetime stemming from gamma-rays and cosmic-ray observations \cite{Massari:2015xea,Ando:2015qda}, which should be considered before claiming any signal of decaying dark matter in the AMS-02 data. We perform a chi-squared goodness of fit and find that a two-component setup with masses of $750$~GeV and $2.3$~TeV, respectively, and a lifetime time of $\sim 10^{26}$~s, yields $\chi^2/d.o.f \sim 0.5$. Such a lifetime is consistent with existing constraints on decaying dark matter. Interestingly, such a lifetime is within reach of current and future gamma-ray telescopes. Searches based on Galactic and extragalactic gamma-ray emission \cite{Ando:2015qda,Massari:2015xea} yield important limits which are within a factor a two from our best-fit points. 

The scenario we advocate here would be testable with continual data taking by the Fermi-LAT mission \cite{Charles:2016pgz} as well as by future gamma-ray observatories such as the Cherenkov Telescope Array \cite{Acharya:2017ttl}, expected to surpass Fermi-LAT's sensitivity on the dark matter lifetime by roughly an order of magnitude \cite{Morselli:2017ree}.
 
\section{Positron Flux} 
\label{sec:indirect}

The AMS-02 collaboration has claimed that the positron flux it measures is incompatible with a background-only hypothesis assuming a standard  diffuse background model \cite{Cavasonza:2016qem}. In order to compute the predicted positron flux, here we combine a background which follows the same assumptions as what reported by the AMS-02 collaboration \cite{Aguilar:2019owu}, and a signal from two-component decaying dark matter. 

The diffuse background is given by the interaction between cosmic rays and the gas in the intergalactic medium and assumed to have a differential flux
\begin{equation}
    \Phi_{back}^{e^+}(E)= c_d\frac{E^2}{\hat{E}^2} \left( \frac{\hat{E}}{E_1}\right)^{\gamma_d}
    \label{backflux}
\end{equation}
where, $c_d=(6.51\pm 0.28)\times 10^{-2}[{\rm m^2\, sr\, s\, GeV}]^{-1}$ is a normalization factor, $\gamma_d=-4.07\pm 0.12$ the spectral index, and $\hat{E} (E)= E + \varphi_{e^+}$, the energy of the particles in the interstellar space, with $\varphi_{e^+}=1.10\pm 0.06$~GeV, all quoted with $2\sigma$ error bars. These values were taken from the supplemental material of Ref.~\cite{Aguilar:2019owu}. $E_1=7$~GeV is a constant chosen to minimize the correlation between the parameters $c_d$ and $\gamma_d$ and $\varphi_{e^+}$ the force field used to account for solar modulation effects. We adopt $c_b=6.9\times 10^{-2} [{\rm m^2\,sr\,s\,GeV}]^{-1}$, $\gamma_b=-3.98$, and $\varphi_{e^+}=1.10$~GeV (notice that the values we choose do not correspond to the central values, but are within 2$\sigma$ of them; our choice is motivated by the fact that with these parameters we find optimal fits to the background plus signal from dark matter; with the central values, our best fit models have a marginally worse chi squared per degree of freedom but continue to provide an excellent fit to the data).

The positron flux for decaying dark matter in the location of the Earth is given by,
\begin{eqnarray}
    \Phi^{e^+}_{\chi} (E) &=& \frac{1}{4\pi b(E)} \frac{\rho_\odot}{m_{\chi}} \Gamma \times \nonumber \\ 
    &\times& \int_E^{m_{\chi}/2}dE_s \sum_{f} BR_f \frac{dN^{e^+}_f}{dE}(E_s)\mathcal{I}(E,E_s) 
    \label{posflux}
\end{eqnarray}
where $b(E)=(E^2 \, \mathrm{GeV})/\tau_\odot$, with $\tau_\odot=5.7 \times 10^{15}$~s, is the energy loss term, $\rho_\odot=0.3$~GeV/cm$^3$, is the dark matter density in the location of the Sun, $\Gamma$ the decay rate,  $dN^{e^+}\hspace{-0.2cm}/dE$ is the spectral energy distribution for positrons in the final state, $\mathcal{I}(E,E_s)$ is the halo function, with $E_s$ the energy of positrons at production in the source, $BR$ is the branching ratio for a given channel $f$. The halo function encodes all the astrophysical uncertainties including the choice for the dark matter density distribution and for the propagation model \cite{Cirelli:2010xx}. Here we adopt a Navarro-Frenk-While profile \cite{Navarro:1995iw} and the ``medium'' propagation model, see Ref.~\cite{Delahaye:2007fr} for details.

\section{Results}
\label{sec:results}

After performing the goodness of fit test with several decay modes with one or two dark matter components, we found that a setup with two different dark matter candidates decaying predominantly into $\tau\tau$ and $\mu\mu$ yields the best-fit to the positron flux. The $\chi^2$ test comparing the predicted and observed positron flux is given by,

\begin{equation}
    \chi^2 = \sum_i \frac{\Phi^{\textit{pred}}_i - \Phi^{obs}_i}{\sigma_i^2}
    \label{chi2}
\end{equation}
where $i$ runs over the energy bins, $\sigma^2_i$ is the sum over squared statistical and systematic uncertainties ($\sigma^2_i=\sigma_{\textit{sys},i}^2 + \sigma_{\textit{stat,i}}^2$) \cite{Aguilar:2019owu}, $\Phi^{\textit{obs}}_i$ the observed flux reported by AMS-02, and $\Phi^{\textit{pred}}_i$ the predicted flux, 

\begin{equation}
    \Phi_{\textit{pred}} = \Phi^{e^+}_{\chi} (E) + \Phi_{back}^{e^+}(E)
\end{equation}
with,
\begin{equation}
    \Phi^{e^+}_{\chi} (E) = \Phi_{\chi_1}^{e^+}(E) + \Phi_{\chi_2}^{e^+}(E)
\end{equation}
where $\Phi_{\chi_1}^{e^+}(E)$ is the expected positron flux from our candidate $\chi_1$, and $\Phi_{\chi_2}^{e^+}(E)$ for the dark matter candidate $\chi_2$. 

\begin{figure}[ht]
    \centering
    \includegraphics[scale=0.45]{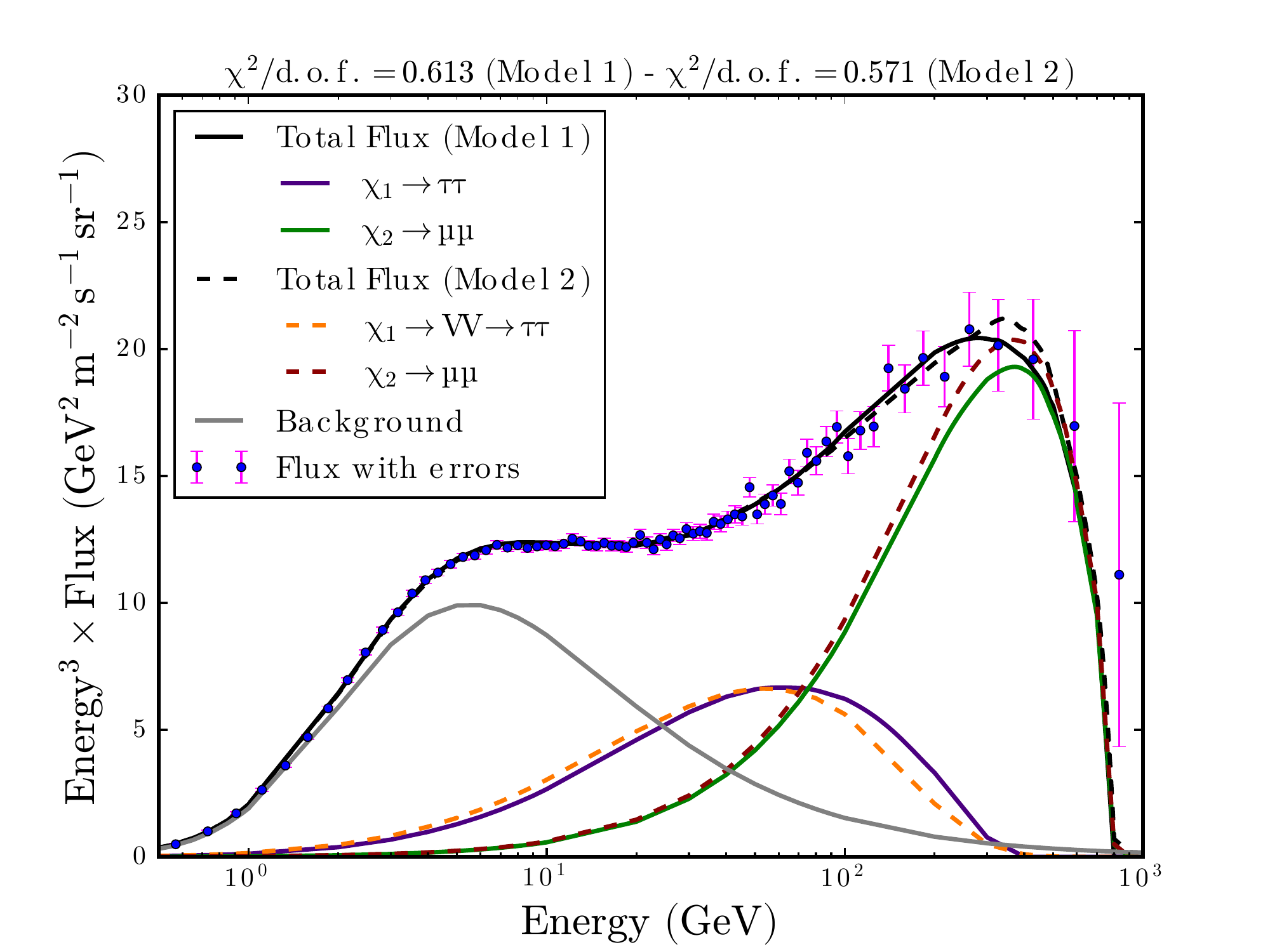}
    \caption{Total positron flux, dark matter plus background, are shown with a solid (dashed) line for model 1 (model 2). Model 1 refers to a two-component decaying dark matter with $\chi_1 \rightarrow \tau\tau$ (purple line) and $\chi_2 \rightarrow \mu\mu$ (green line). Model 2 accounts for the $\chi_1 \rightarrow VV \rightarrow 4\tau$ (orange line) and $\chi_2 \rightarrow \mu\mu$ (red curve). The data is provided by AMS collaboration and the background is shown with a grey line \cite{Aguilar:2019owu}.}
    \label{fig:flux}
\end{figure}

We emphasize that in order to compute the flux given in Eq.~\ref{posflux}, we use the PPPC4DMID package \cite{Cirelli:2010xx} with the ``medium'' propagation model as defined in \cite{Cirelli:2010xx}, and the diffuse background in Eq.~\ref{backflux}. The combined fluxes (black line), including the background given by Eq.~\ref{backflux} (grey line), the dark matter flux with contributions from both dark matter components $\chi_1$ (purple line) and $\chi_2$ (green line) are showed in Fig.~\ref{fig:flux}, here we choose for $\chi_1$, $m_{\chi_1}=750$~GeV with decay rate $\Gamma_{\chi_1 \rightarrow \tau\tau}=7.00\times10^{-27}$~s$^{-1}$, which gives a lifetime of $\tau_{\chi_1}=1.43\times10^{26}$~s, and for $\chi_2$ , a mass $m_{\chi_2}=2300$~GeV with decay rate $\Gamma_{\chi_2 \rightarrow \mu\mu}=5.46\times10^{-27}$~s$^{-1}$, which results into a lifetime of $\tau_{\chi_2}= 1.83\times10^{26}$~s. We emphasize that we assumed that each dark matter component account for $50\%$ of the dark matter local density, which results into a normalization in fluxes by a factor of two. The resulting $\chi^2$, given by Eq.(\ref{chi2}), gives a $\chi^2/d.o.f.=42.93/70 = 0.613$, hence a remarkably  good fit to the data. 

We have also investigated the possibility of replacing the decay into $\tau\tau$ by a secluded decay where $\chi_1 \rightarrow VV\rightarrow 4\tau$. It will be clear later on why we assessed this channel; in this case, the best fit is achieved for  $m_{\chi_1} = 1150$~GeV with a lifetime $\tau_{\chi_1} = 1.49 \times 10^{26}$~s, and $m_{\chi_2} = 2300$~GeV with a lifetime $\tau_{\chi_2} = 1.73 \times 10^{26}$~s. The fluxes generated from this scenario are shown with dashed curves in Fig.~\ref{fig:flux}, where we conclude that such a setup also provides a good fit to data.

\begin{figure}[ht]
    \centering
    \includegraphics[scale=0.45]{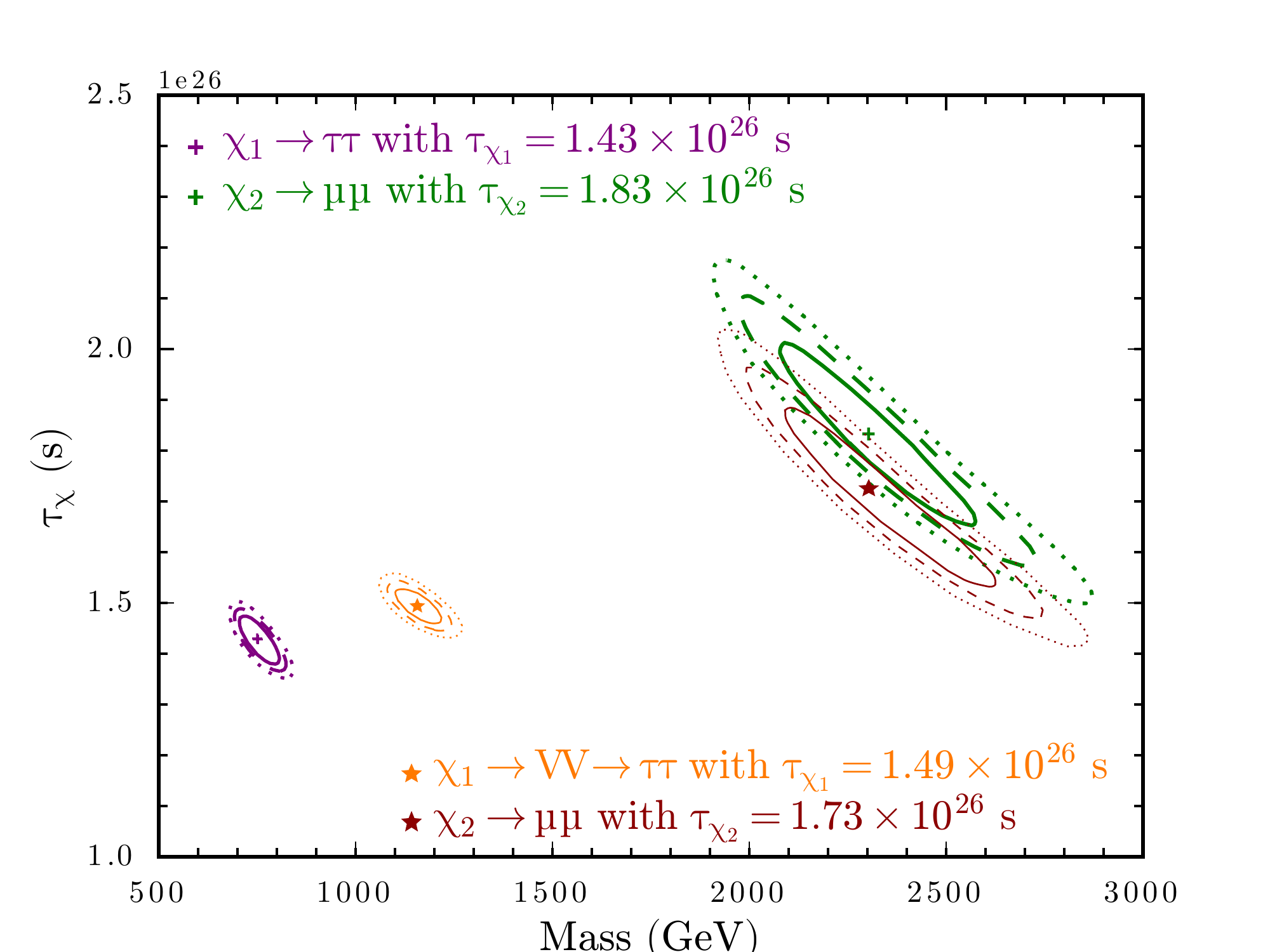}
    \caption{Best fit regions for lifetime for both dark matter candidates, for Model 1 we have $\chi_1$ (purple contours) and $\chi_2$ (green contours), while for Model 2 we have $\chi_1$ (orange contours) and $\chi_2$ (red contours), including 1$\sigma$, 2$\sigma$ and 3$\sigma$ regions, respectively. The cross (star) marks the best-fit point with central values given in the legend for Model 1 (Model 2).}
    \label{fig:cs}
\end{figure}

In Fig.~\ref{fig:cs}, we display the best-fit contours in terms of the {\it lifetime vs mass} of the dark matter particles, with $1\sigma$ (continuous line), $2\sigma$ (dashed line) and $3\sigma$ (dotted line) for each decaying dark matter scenario studied. We have kept the same color scheme as Fig.~\ref{fig:flux}.

In the first case, we notice that the best-fit is found for  $\tau_{\chi_1}=1.43\times 10^{26}$~s, $\tau_{\chi_2}=1.83 \times 10^{26}$~s, with  $m_{\chi_1}=750$~GeV, $m_{\chi_2}=2.3$~TeV. 
It is noticeable that with this simple two-component decaying dark matter we can find such a good fit to the data. It is important to point out there are stringent limits rising from gamma-rays observations. The best-fit point found for the $\mu\mu$ decay is consistent with gamma-ray observations but it is just a factor a two below from existing bounds. Thus, it can be tested in the near future. The lifetime we inferred for the $\tau\tau$ decay that offers a best-fit to data is, however, disfavored by gamma-ray searches for decaying dark matter in the Milky Way halo that impose  $\tau_{\chi_1} > 3.6\times10^{26}$~s \cite{Massari:2015xea}. However, there are large uncertainties on the Inverse Compton scattering contribution from propagation and energy losses of charged particles \cite{Massari:2015xea}, and the fact that the AMS signal is local, i.e. originating from within $1-2$~kpc at most, while the gamma ray constraints stem from much more distant regions. Therefore, it is plausible to assume that this bound is subject to uncertainties, at least larger than a factor of a few, which are sufficient to make our best-fit point for the decay into $\tau\tau$ consistent with the current limits from gamma-rays. 

Furthermore, setting aside these uncertainties and concentrating on Model 2, where the decay into $\tau\tau$ is replaced by a secluded ($\chi_1 \rightarrow VV \rightarrow 4\tau$) scenario we find that $m_{\chi_1} = 1150$~GeV with a lifetime $\tau_{\chi_1} = 1.49 \times 10^{26}$~s, and $m_{\chi_2} = 2300$~GeV with a lifetime $\tau_{\chi_2} = 1.73 \times 10^{26}$~s, yields $\chi^2/d.o.f.=0.571$.  Since the limit stemming from gamma-ray observations on such a secluded decay mode is much weaker, this second possibility is entirely open \cite{Massari:2015xea}.

We note that it is conceivable to have other configurations with similarly good fit quality where the dark matter candidates have densities different from this 50\%-50\% setup. Since gamma-ray constraints scale as $\rho_i/\tau_i$, a smaller (larger) relative dark matter abundance would be allowed for a shorter (respectively longer) lifetime. We find nonetheless that a 50\% 50\% scenario generally provides a close-to-optimal spectral fit for the AMS-02 data.

Having in mind that we can fit the data with this two-component dark matter setup one may wonder if this is theoretically feasible. We address this issue presenting a sketch of three different possible models featuring real scalars, Majoron fields, and vector bosons which can potentially accommodate our scenario.

\section{Sketches of Concrete Models}

\subsection{Two real scalars}\label{sec:mod1}
We postulate two real scalar fields $S_1$ and $S_2$ charged under two distinct $Z_2$ symmetries, $Z^1_2\times Z_2^2$, subject to a potential such that the fields do not acquire a non-zero vacuum expectation value (v.e.v.) upon spontaneous breaking of the SU(2) gauge symmetry. The Lagrangian of the model includes, in addition to the Standard Model (SM) Lagrangian, the following renormalizable terms:
\begin{equation}
    {\cal L}\supset m_1^2S_1^2+m_2^2S_2^2+\lambda_0 S_1^2S_2^2+\lambda_1H^\dagger H S_1^2+\lambda_2H^\dagger H S_2^2.
\end{equation}
The parameters $m_i$ are fixed at the masses $750$~GeV and $2.3$~TeV. The couplings $\lambda_i$ are chosen so as to induce the correct thermal relic density of dark matter for both species. 

Notice that the $S_i$ are absolutely stable here. If such discrete symmetries originate from global symmetries at larger scales, the long-lived decays are induced by the only Planck-suppressed operators we postulate to exist, according to the analysis of Ref.~\cite{Mambrini:2015sia},
\begin{equation}
    {\cal L}_{\rm Pl}\supset \frac{\kappa_i}{M_{\rm Pl}}\partial_\mu S_i \bar f_i\gamma^\mu\left(1\pm\gamma^5\right)f_i,
\end{equation}
with $f_1=\tau$, $f_2=\mu$; these operators induce a decay width 
\begin{equation}
    \Gamma=\frac{\kappa_i^2}{2\pi}\frac{m_f^2 m_{i}}{M^2_{\rm Pl}}.
\end{equation}
To match the lifetimes in accordance with the observed excess we need $\kappa_1\sim8.4\times 10^{-9}$ and $\kappa_2\sim4.2\times 10^{-9}$.

The model can be slightly changed by postulating that the $Z_2^1$ symmetry results from spontaneous symmetry breaking of the gauged tau-lepton number U(1)$_\tau$ symmetry with a massive gauge boson $Z^{\prime}$ of mass $m_{Z^\prime}\ll m_{1}$; we then postulate that $\kappa_1=0$, but that $\kappa_1^\prime\neq$ zero, where $\kappa_1^\prime$ is the coefficient of a different dimension-5 Planck-suppressed operator,
\begin{equation}\label{eq:dim5op2}
    \frac{\kappa^\prime_1}{M_{\rm Pl}}S_1Z^{\prime\mu\nu}Z^{\prime}_{\mu\nu};
\end{equation}
The decay width induced by the dimension-five operator in Eq.~(\ref{eq:dim5op2}), neglecting the mass of the $Z^\prime$, is
\begin{equation}
    \Gamma\simeq\frac{\kappa^{\prime 2}_1m_1^3}{4\pi M_{\rm Pl}^2}.
\end{equation} now with $m_1=1150$ GeV, which reproduces the current lifetime for $\kappa^{\prime}\simeq 1.5\times 10^{-11}$.

\subsection{Majorons}\label{sec:mod2}

A second possibility consists of two massive Majorons, along the lines of the model described in Ref.~\cite{Garcia-Cely:2017oco}. Here we postulate the masses to match the ones that fit the data, which implies a see-saw scale $f\gtrsim 10^{12}$ GeV. A decay width with the desired flavor structure can be guaranteed by an appropriate flavor matrix $K$, following the notation of Ref.~\cite{Garcia-Cely:2017oco}, with the flavor-dependent diagram in Fig.~1, (b) compensating for the flavor-diagonal diagram (a). 

In this model, it is inevitable to have a large, tree-level decay rate into neutrinos, and, also, one-loop decay to quarks, and two-loop level decay to gauge bosons and SM Higgs. In any case, a Majoron model augmented with a broken $U(1)_{L_{\mu,\tau}}$ would certainly allow for the right decay pattern. In the presence of a spontaneously broken $U(1)_{L_{\tau}}$ it is also possible to enhance the decay of one of the Majorons predominantly into the corresponding gauge boson, which  would then decay to $\tau\tau$.

As far as production of the Majoron is concerned, both freeze-out and freeze-in should be possible, according to the discussion in Sec.III of Ref.~\cite{Garcia-Cely:2017oco}.
\subsection{Vector bosons}\label{sec:mod3}
Here we postulate the following additional (spontaneously broken) gauge groups:
\begin{equation}
    {\cal G}_{\rm SM}\times \Big({\rm U}_1(1)\times {\rm U}_{L_\tau}(1)\Big)\times \Big({\rm U}_2(1)\times {\rm U}_{L_\mu}(1)\Big) ,
\end{equation}
where the ${\rm U}_{L_l}(1)$  correspond to gauged muon ($l=\mu$) and tau $(l=\tau$) lepton numbers. All of the U(1)'s are assumed to be spontaneously broken with the corresponding gauge bosons having mass $m_{Z_1}=750$ GeV, $m_{Z_2}=2300$ GeV, and $m_{Z_\mu},\ m_{Z_\tau}\ll m_{Z_1}$. The gauge bosons fields are postulated to have the following kinetic mixing structure:
\begin{equation}\label{eq:kinmix}
    \epsilon_1 (F_1)_{\mu\nu}(F_\tau)^{\mu\nu}+\epsilon_2 (F_2)_{\mu\nu}(F_\mu)^{\mu\nu},
\end{equation}
with all other possible kinetic mixing (including with Standard Model photons) vanishing. The correct abundance of $Z_i$ is ensured by resonant freeze-in in the early universe (the $Z_\mu$ and $Z_\tau$ at early times are relativistic and in equilibrium with the SM thermal bath; resonant production kicks in when the thermal masses of the $Z_f$'s match the masses of the $Z_i$).

The decay width induced by the kinetic mixing reads:
\begin{equation}
    \Gamma_i\simeq \frac{\varepsilon_i^2m_{Z_i}}{8\pi},
\end{equation}
where $\varepsilon_i=\epsilon_i g_i$ and it produces the desired decays as long as $\varepsilon_1\sim 7\times 10^{-27}$ and $\varepsilon_2\sim 1.1\times 10^{-26}$. Notice that while these values of $\varepsilon_i$ are very small, they are technically natural (in the usual sense that the limit $\varepsilon_i\to0$ increases the symmetries of the model's Lagrangian).

To accommodate the $Z_1\to Z_{\tau}Z_\tau$ we need either to postulate that the gauge symmetries to be non-Abelian (in which case the kinetic mixing in Eq.~(\ref{eq:kinmix}) is not allowed, and one would need to resort to a higher-dimensional, Planck suppressed operator to mediate the decay), or, more simply, that the gauge group above be reduced to
\begin{equation}
{\cal G}_{\rm SM}\times \Big({\rm U}_2(1)\times {\rm U}_{L_\mu}(1)\Big) \times \Big({\rm U}_1(1)\Big),
\end{equation}
and the particle content augmented by a scalar $S_1$ with the following interaction with the $Z_1$:
\begin{equation}
   g Z_1^\mu S_1\partial_\mu S_1,
\end{equation}
with the $S_1$ decaying as in the first model discussed above to $\tau^+\tau^-$ via a Planck-suppressed dimension-5 operator.
\section{Discussion} 
\label{sec:con}

Over the past decades several experiments reported an excess in the flux of cosmic-ray positrons for energies above $10$~GeV. The new data release from the AMS-02 collaboration has offered new important information that begs for a new theoretical interpretation. The smaller error bars, the absence of a flat spectrum for energies above $150$~GeV, a peak at 300 GeV, and a cut-off at $\sim 700$~GeV, present a challenge and an opportunity to explain the extra positrons from new physics. Adopting the background recommended by the AMS-02 collaboration, we found a very good fit to the data with a two-component decaying dark matter. Our best-fit yields $\chi^2/d.o.f \sim 0.5$ for a two-component decaying dark matter with a lifetime of the order $\tau_{\chi}\sim 10^{26}$~s for $m_{\chi}=750$~GeV in the $\tau\tau$ channel, and for $m_{\chi}=2.3$~TeV in the $\mu\mu$, which is compatible with the existing limits. One could also find a good-fit by replacing the $\tau\tau$ channel by a secluded decay where a dark matter particle decays into a pair of bosons that later decay producing $4\tau$. This second, admittedly more contrived possibility is marginally favored by weaker constraints from gamma-ray observations. We also note that a purely secluded two-component decaying dark matter where the decay into $\mu\mu$ is also traded for a decay into $4\mu$ via dark boson decays, would also provide a good fit to the data. We checked that in a pure secluded two-component dark matter decay one would get a best fit $\chi^2/d.o.f =0.571$ for $m_{\chi_1}(m_{\chi_2})=1200 (4000)$~GeV, with $\tau_{\chi_1,\chi_2}= 1.6 \times 10^{26}$~s, which is in agreement with current bounds. Thus, either way, our two-component decaying dark matter setup is capable of reproducing the data while circumventing existing bounds on the dark matter lifetime from gamma-ray observations.  A possible way to distinguish a model from the other would be via an independent observation capable to pinning down the dark matter mass, since secluded decays favor larger masses.

We have shown that a two-component decaying dark matter is theoretically appealing by describing how one could embed it in three different dark matter models. Given the precise measurements reported by AMS-02, assuming the validity of the background model adopted by the AMS-02 collaboration, we find circumstantial evidence for a possible signal of two-component decaying dark matter. Intriguingly, our best-fit points will be probed by future gamma-ray telescopes.
\section*{Acknowledgments}
SP is partly supported by the U.S.\ Department of Energy grant number de-sc0010107. FSQ acknowledges support from MEC and ICTP-SAIFR FAPESP grant 2016/01343-7. FSQ and CS are supported by MEC and UFRN. SP gratefully acknowledges input and feedback from many members of the SCIPP theory group, including  Nicolas Fernandez, Akshay Ghalsasi,   Logan Morrison and, especially, Wolfgang Altmannshofer, Stefania Gori, Hiren Patel and Bibhushan Shakya for detailed feedback on this manuscript.

\bibliographystyle{JHEPfixed}
\bibliography{darkmatter}

\end{document}